\documentclass[12pt,preprint]{aastex}

\shorttitle{Sunyaev-Zel'dovich Angular Power Spectrum}
\shortauthors{Bode, Ostriker, \& Cen}

\begin{document}
\title{ Calibration of Nonthermal Pressure in Global Dark Matter
          Simulations of Clusters of Galaxies}
\author{ Paul Bode, Jeremiah P. Ostriker, and Renyue Cen }
\affil{Department of Astrophysical Sciences, Princeton University,
Princeton, NJ 08544}
\and
\author{Hy Trac}
\affil{Department of Physics, Carnegie Mellon University, 
Pittsburgh, PA 15213}

\begin{abstract}
We present a new method for incorporating nonthermal pressure
from bulk motions of gas
into an analytic model of the intracluster medium in clusters of
galaxies, which is
based on a polytropic equation of
state and hydrostatic equilibrium inside gravitational potential wells
drawn from cosmological dark matter simulations. 
The pressure is allowed to have thermal and nonthermal components
with different radial distributions;
the overall level of  nonthermal support is based on the dynamical
state of the halo, such that it is lower in more relaxed clusters.
This level is normalized by comparison to 
pressure profiles derived from X-ray observations,
and to a high resolution hydrodynamical simulation. 
The nonthermal pressure fraction measured at $r_{500}$
is typically in the range 10-20\%, increasing with cluster
mass and with redshift.
The resulting model cluster properties are in accord with
Sunyaev-Zel'dovich (SZ) effect observations of clusters.
Inclusion of nonthermal pressure reduces the 
expected angular power spectrum of SZ fluctuations in the microwave
sky by 24\%.
\end{abstract}


\section{Introduction} \label{sec:intro}

\defcitealias{ShawNBL10}{SNBL}
\defcitealias{BodeOV09}{BOV09}

The millimeter and
microwave sky has recently been mapped on small angular scales
to much higher precision than was previously possible, 
most notably by 
the South Pole Telescope
\citep[SPT][]{LuekerSPT10,ShirokoffSPT11,ReichardtSPT11z} and
the Atacama Cosmology Telescope \citep[ACT][]{FowlerACT11,DasACT11}.
These telescopes, as well as the PLANCK satellite, will
soon provide further results with increased frequency and sky coverage.
At these scales (multipoles $\ell \gtrsim 3000$) the
primordial cosmic microwave background is strongly damped, 
and other components dominate. 
After accounting for point sources, such as
dusty star forming galaxies and radio galaxies
\citep{HallSPT10,VieiraSPT10,MarriageACT11a},
the main contributor to
microwave anisotropies is then the Sunyaev-Zel'dovich (SZ)
effect from galaxy clusters, groups, and the intergalactic medium.
SZ measurements
can give strong constraints on cosmological parameters
\citep[e.g.][]{KomatsuWMAP11,DasACT11,DunkleyACT11,SehgalACT11,ReichardtSPT11z},
particularly the normalization of the matter power spectrum,
conventionally given as $\sigma_8$,
the amplitude of fluctuations on an 8$h^{-1}$Mpc scale.


However, these constraints rely on the ability to predict
the SZ signal for a given set of cosmological parameters.
This requires knowledge of the state of the intracluster medium (ICM)
in groups and clusters out to high redshift.
Many physical processes shape
the ICM, including shocks from accretion and merging,
star formation, supernovae, and active galactic nuclei (AGN),
among others.  Our understanding of these processes and their
effect on the ICM over time is still evolving.
Measurements of how the relations among X-ray and SZ observables
(and their relation to mass) evolve with redshift are ongoing;
current observational constraints at higher $z$ leave much
room for improvement \citep{ReichertBFM11}.
On the theoretical side,
lacking analytic solutions for gravitational collapse and many aspects
of the behavior of baryons, recourse is taken to numerical simulations.
Full hydrodynamic simulations with all the required physical inputs
are expensive to run and involve many details of subgrid physics;
different numerical schemes can give varying results
\citep{Agertz07,MitchellMBTC09,VazzaDRBGKP11,SijackiVKSH11z}.
Thus, global predictions of the expected SZ signal
are currently highly uncertain.

With this in mind, we have developed a method for determining
the ICM distribution in a given gravitational potential well
\citep{OstrikerBB05,BodeOWS07,BodeOV09,TracBO11}.
This makes it possible to take the output of a 
large volume cosmological simulation with
only dark matter (DM) and, by populating the halos with gas,
produce predictions for the SZ signal over a large area of the sky.
To date, 
this model has assumed that the gas is in hydrostatic equilibrium (HE)
and that it has a polytropic equation of state.
Levels of star formation and feedback energy are set so
that the model matches current X-ray observations.
Based on this, the ``Standard'' model of
\citet[][hereafter BOV09]{BodeOV09} was used to produce
an all-sky SZ map \citep{SehgalBDHHLOT10};  this yields a
template for the angular power spectrum which can be
compared with microwave observations.
SPT \citep{LuekerSPT10,ShirokoffSPT11,ReichardtSPT11z}
and ACT \citep{FowlerACT11,DunkleyACT11}
both found that the observed power is lower (by up to a factor of 2)
than expected from this template.
A possible explanation for this discrepancy is that the
\citetalias{BodeOV09} model overestimates the thermal
pressure in the outer regions of clusters.
X-ray scalings, to which the model is normalized, are generally
not measured beyond $r_{500}$
(the radius within which the cluster density is 500 times the 
critical value) because this region dominates the signal,
whereas the SZ angular power spectrum
is significantly affected by the regions outside this radius.

In particular, bulk motions of gas and turbulent
pressure support will occur, making the
assumption of HE incorrect.  This would be more likely
in the outer regions of clusters, where the gas has been
accreted more recently, and in younger, less dynamically relaxed  clusters.
There is considerable support from simulations for
this picture, with the temperature dropping below that
which would be predicted by HE
\citep{LauKN09,BurnsSO10,VazzaBGBB11,BattagliaBPSI}.
Though difficult to obtain, there is some observational
evidence that nonthermal sources of pressure are present
in cluster outskirts.
Drops in the average temperature with radius, leading to
a flattening in the radial profile of entropy,
have been seen
\citep{MahdaviHBH08,GeorgeFSYR09,BautzMSAMea09,HoshinoHSAYea10,UrbanWSAB11,AkamatsuIOSTO11,MorandiLSGCPA11z},
although this is not always the case
\citep{ReiprichHZSea09,EttoriB09,SimionescuAMWea11,HumphreyBBFGM11}.
Gas clumping  \citep{SimionescuAMWea11,UrbanWSAB11},
deviations from symmetry \citep{EckertVEMea11z,WalkerFSGT12z},
and the fact that backgrounds dominate \citep{EckertMGR11}
complicate interpretation of these observations.

\citet{TracBO11} included nonthermal pressure
in the polytropic model, and found the resulting
SZ template significantly reduced the tension with current SZ measurements.
However, this had the limitation of assuming
that the fraction of  nonthermal pressure was the
same (20\%) at all radii, and the same for all cluster
masses and redshifts, when in fact physical modeling
indicates that the nonthermal fraction should be larger
in the outer parts of clusters, and should also be
larger in recently formed clusters (i.e.\ increase with
increasing redshift).

The assumption of a polytropic equation of state means
the gas density is given by
$\rho = \rho_0 \theta^\frac{1}{\Gamma-1}$
and the pressure by
$P = P_0 \theta^\frac{\Gamma}{\Gamma-1}$,
where the polytropic variable $\theta$ is a function
of the potential, the central density $\rho_0$,
and the central pressure $P_0$
(see  Eqn.\,\ref{eqn:3dtheta}).
Based on hydrodynamical simulations of cluster formation,
\citet{OstrikerBB05} assumed $P$ represented  thermal pressure
only, and argued that the appropriate polytropic index was $\Gamma=1.2$.
\citet[][hereafter SNBL]{ShawNBL10}, based on more recent
simulations, 
instead argued that it is the {\it sum} of
thermal and turbulent pressures which follows the polytropic relation
(but still with $\Gamma=1.2$).
In this case the fraction of nonthermal pressure varies within
a cluster, increasing with radius, and also increases with redshift.
The simulations of
\citet{BattagliaBPSI} give support for a cluster mass dependence as well.

In this paper we modify the model of \citetalias{BodeOV09}
to include a treatment of nonthermal pressure similar to that presented in 
\citetalias{ShawNBL10}.  Our approach will differ from 
\citetalias{ShawNBL10} in that it will be
based on the dynamical state of each individual halo,
such that less relaxed halos have a higher level of turbulent
pressure.  The relation between the level of virialization
and the fraction of turbulent pressure is calibrated by
comparison to cosmological hydrodynamic simulations 
(in particular a high resolution run using the ENZO code)
and the pressure profile derived from X-ray observations.

Sec.\,\ref{sec:model} summarizes the polytropic model.
Sec.\,\ref{sec:nonthp} presents the new method for
including turbulent pressure, which includes a free
parameter that is set by normalizing to observed and
simulated clusters.  
Note the model parameters are set by comparison
to X-ray data only.
The model, applied to a large set of halos, is then compared
to SZ observations in Sec.\,\ref{sec:discuss}.

\section{Polytropic Model} \label{sec:model}

The goal behind this work is, given a particular cosmology,
to produce a simulated map of the microwave sky.
The strategy is to first run a pure dark matter simulation of
a large volume, saving the mass distribution along the
past light cone.  The group and cluster sized halos in the light
cone are identified, and each in turn is populated with
gas;  thus the thermal and kinetic SZ effect arising from
clusters can be calculated.  In this paper we will use halo
samples drawn from a light cone output of a simulated cube
1000 $h^{-1}$Mpc on a side (see \citetalias{BodeOV09} for more details)
with cosmological parameters
($\Omega_b, \Omega_m, \Omega_{\Lambda}, h, n_s, \sigma_8$) =
(0.044, 0.264, 0.736, 0.71, 0.96, 0.80), consistent with
the WMAP 7-year results \citep{KomatsuWMAP11}.

Adding gas to halos is done with the method of \citetalias{BodeOV09}.
Gas is placed in
the DM gravitational potential $\phi$
such that it is in hydrostatic equilibrium
and has a polytropic equation of state.
The polytropic variable $\theta$ is
\begin{equation}
\theta \equiv 1 + \frac{\Gamma-1}{\Gamma}
  \frac{\rho_0}{P_0}\left( \phi_0-\phi \right)   \;,
\label{eqn:3dtheta}
\end{equation}
where $\phi_0$, $\rho_0$, and $P_0$ are the central values
at radius $r=0$.
\citetalias{BodeOV09} assumed that $P_0$ was due to thermal pressure
alone.  
Here we will instead follow \citetalias{ShawNBL10},
so that it is the sum of
thermal and turbulent pressures which follows the polytropic relation.
The turbulent component will have the same compressibility as the
thermal component, so the polytropic index remains $\Gamma=1.2$.
Thus the total pressure is the sum of thermal $P_{th}$ and
turbulent $P_{nth}$ components:
\begin{eqnarray} \label{eqn:press}
P = P_{th}+P_{nth}
        = P_0 \theta^\frac{\Gamma}{\Gamma-1}
\hspace{1cm} , \\
f_{nth} = P_{nth}/(P_{th}+P_{nth})
\hspace{1cm} , \\
P_{th} = (1-f_{nth}) P_0 \theta^\frac{\Gamma}{\Gamma-1}
\hspace{1cm} .
\end{eqnarray}
The density profile is unchanged from \citetalias{BodeOV09}, 
but the temperature is altered by a factor $(1-f_{nth})$.
$f_{nth}$ varies with radius,
so the effective polytropic
index, $\Gamma_{eff}=d\log P_{th}/d\log \rho$ is not
constant, but rather increases with radius.
We will let
$f_{nth}$ also vary with cluster mass
and redshift (see Sec.\,\ref{sec:nonthp}).

Two constraints are used to fix the central values
$\rho_0$ and $P_0$ \citepalias[see][]{BodeOV09}.
First, the total pressure at the outer radius must match
the surface pressure $P_s$ (calculated from the DM in a
buffer region at $r_{vir}$). 
Second, energy must be conserved.
Assume the gas initially followed the DM density distribution
inside the virial radius $r_{vir}$, with the same specific energy.
Some of the gas (that with the lowest entropy)
is removed and placed into stars.  The remaining
gas has an initial energy $E_g$;  this gas is now rearranged to obey
the polytropic equation of state.
The energy of the gas in this final arrangement is
$E_f = \frac{3}{2} \int  P dV$
(the integral is over the cluster volume, out to
the outermost extent of the rearranged
gas, $r_f$; note $r_f$ can be larger than $r_{vir}$).
This energy must
equal the initial value, modified by various processes
(see \citetalias{BodeOV09} for details):
\begin{equation} \label{eqn:3dfe}
E_f = E_g + \Delta E_P + \epsilon_D\left| E_D\right| + \epsilon_FM_*c^2
\hspace{1cm} .
\end{equation}
The first modification to the initial gas energy
is a change due to expansion (or contraction) of the gas,
$\Delta E_P =(4\pi/3)(r_{vir}^3-r_f^3)P_s$, which accounts
for mechanical work done against the surface pressure.
The next term is a dynamical
energy input, needed because during cluster formation
energy is transferred from DM to gas by
dynamical processes \citep{McCarthyBBVPTBLF07}.  
The binding energy $E_D$ of the DM is found,
and 5\% ($\epsilon_D=0.05$) is added to the gas;
this value provides a good match to the gas fractions seen in
``adiabatic'' (no radiative cooling or star formation)
hydrodynamic simulations \citep{CrainEFJMNP07,YoungTSP11}.

The final term in Eqn.\,\ref{eqn:3dfe}
is feedback from star and black hole
formation, where $M_*$ is the mass of stars inside the virial radius
The mass in stars at $z=0$ is set to
\begin{equation}
M_*/M_{vir} = A_*\left( 5.0\times 10^{13} M_\odot/ M_{500}
\right)^{\alpha_*}
\;,
\end{equation}
where $M_{500}$ and $M_{vir}$ are the total mass inside
$r_{500}$ and $r_{vir}$, respectively.
From the infrared properties of clusters, \citet{LinMS03} found
$A_*=0.026$ and $\alpha_*=0.26$ when measuring the stellar mass
inside $r_{500}$.  
For nearby, low-mass galaxy clusters, \citet{BaloghMBEBLT11}
find stellar fractions in agreement with \citet{LinMS03}.
Note we use these parameters to give $M_*$
for the entire cluster out to the virial radius
(not just inside $r_{500}$);
\citet{Andreon10} found that this choice
gives a model that matches the observed stellar fraction within
$r_{200}$.
This stellar fraction differs (particularly at lower masses)
from that found by \citet{Giodiniea09}, who obtained
$A_*=0.05$ and $\alpha_*=0.37$ (again measuring inside $r_{500}$).
However,
we find that with the \citet{LinMS03} values the model clusters
give a better match to observed electron pressure profiles
(see Sec.\,\ref{sec:nonthp}).
The redshift evolution of stellar mass, including gas recycling, is 
handled in the manner described in \citetalias{BodeOV09}.
\citet{LinSEVMK12} find little evolution in the stellar fraction
relative to \citet{LinMS03} out to $z\sim 0.6$.
At higher redshifts ($\sim$0.2--1), and
measuring out to 200 times the mean density,
\citet{Leauthaud11x} find total stellar fractions lower than what
we are using here \citep[see also][]{Leauthaud11y}; 
adjusting for intra-cluster light would at least
partially reduce the difference.

Feedback energy, from both supernovae and AGN, is
parametrized as $\epsilon_{F}M_*c^2$.
This will be set to match observations (Sec.\,\ref{sec:feedback});
we find $\epsilon_F=8\times 10^{-6}$.
If the mass in black holes is $\sim 10^{-3}M_*$, this means
roughly $6\times 10^{-3}$ of the black hole rest mass is deposited
in the surrounding gas (ignoring the supernova contribution).
A similar value of
$5\times 10^{-3}$ is needed in simulations to produce the correct
relationship between black hole mass and stellar velocity dispersion
\citep{DiMatteoSH05}.
If 15\% of the baryons are turned into stars, then this level
of feedback adds 1 keV per particle initially inside the virial radius;
this is actually lower than the value of $2.62\pm 0.85$ keV per particle
\citet{ChaudhuriNM12z} derived from X-ray data.

It is possible to include 
relativistic pressure, e.g.\ cosmic rays and tangled magnetic fields,
under the assumption
that relativistic contribution is proportional to the
non-relativistic pressure, i.e.\ 
$P_{rel} = \delta_{rel}(P_{th}+P_{nth})$
\citep[see][]{BodeOV09,TracBO11}.
However, we will not do so
(in effect setting $\delta_{rel}=0$), because including this effect
did not improve the fit to observed clusters.

\section{Incorporating nonthermal pressure} \label{sec:nonthp}
\subsection{Nonthermal pressure from turbulence and bulk
motions} \label{sec:subnonthp}
Based on adaptive mesh refinement hydrodynamic simulations 
of cluster formation
(which include star formation and feedback), \citetalias{ShawNBL10}
found that the radial profile of the nonthermal pressure fraction
follows  a power law in radius:
\begin{equation}
f_{nth} = \alpha \left( \frac{r}{r_{500}} \right)^{n_{nth}} \;\;,
\label{eq:nonthermalpress}
\end{equation}
with $n_{nth}=0.80\pm0.25$. 
Examining their simulations out to $z=1$,
\citetalias{ShawNBL10} derived a redshift-dependent normalization
\begin{equation}
\alpha(z) = {\rm MIN} [ 
\alpha_0 (1+z)^{\beta}, \;
(4^{-n_{nth}}\alpha_0^{-1} - 1)\tanh (\beta z) + 1 ] \;\;,
\label{eq:zevo}
\end{equation}
where $\alpha_0=0.18\pm0.06$ and $\beta=0.5$.
\citet{BattagliaBPSI} also used simulations
to constrain $f_{nth}$, with clusters out to $z=1.5$ drawn from
a set of smoothed particle hydrodynamics cosmological boxes;
star formation and AGN feedback were included.
They saw redshift evolution similar to
\citetalias{ShawNBL10}, but also a weak mass dependence;
for $z\lesssim1$ this
can be included by modifying $\alpha$ to be
\begin{equation}
\alpha(z,M) = \alpha_0  (1+z)^{\beta}
  \left( \frac{M_{200}}{3\times 10^{14}M_\odot)} \right)^{0.2} 
\;\;.
\label{eq:mzevo}
\end{equation}
These results seem to be robust with regard to the details
of the gas physics, as both \citetalias{ShawNBL10} and
\citet{BattagliaBPSI} obtain 
similar values when not including gas cooling and star formation.

This can easily be incorporated into the method of \citetalias{BodeOV09}.
One possibility is to simply use 
Eqn.\,\ref{eq:zevo} or \ref{eq:mzevo} for $\alpha$.
However, it is seen in simulations that
relaxed clusters tend to have less turbulent pressure
support than unrelaxed ones 
\citep{LauKN09,VazzaBGBB11,PaulIMBM11,HallmanJeltema11,NelsonRSN11z}.
Thus, rather than giving all halos at a given mass and redshift
the same value for nonthermal fraction, 
it would be more accurate to base $\alpha$
on the dynamical state of the halo. 
We will do this as follows in the next section.

\subsection{Setting the level of nonthermal pressure} \label{sec:normnonthp}
The mean nonthermal energy fraction in the gas is given by
$\bar{f} =  \int P_{nth} dV / \int  P dV$,
with the integrals over the cluster volume out to the virial radius.
In a situation where the density and temperature distribution
of the gas is known (e.g.\ from an observed cluster or a hydrodynamical
simulation), 
the fraction of energy due to 
turbulent nonthermal pressure can be computed directly as
\begin{equation}
\bar{f}  =
\frac{ \int \frac{1}{3}\rho\sigma^2dV }
{ \int \left( \frac{\rho kT}{\mu m_p} + \frac{1}{3}\rho\sigma^2 \right) dV }
\;,
\end{equation}
where $\sigma^2$ is the velocity dispersion,
$k$ is the Boltzmann constant, and $\mu m_p$ is the
mean mass of the ICM particles.

When determining the polytropic gas distribution for a DM-only halo,
the total pressure $P$ is found 
before determining the turbulent and thermal components separately.
For a given cluster we can thus compute
a measure $f^\prime_{pol}$ proportional to the nonthermal energy fraction
$\bar{f}_{pol}$
when assuming the form of Eqn.\,\ref{eq:nonthermalpress}:
\begin{equation}
\bar{f}_{pol} = 
\frac{ \int \alpha (r/r_{500})^{n_{nth}} P dV } { \int P dV }
= \alpha f^\prime_{pol}   \;.
\end{equation}

We wish to base the value of
$\bar{f}  = \alpha f^\prime_{pol}$
on the properties of the DM halo;
thus a measure of how relaxed or disturbed a halo may be
is required.
The virial theorem states
$\ddot{I}/2 = 2K + W - E_S$, 
where $K$ and $W$ are the kinetic and potential energies,
$I$ is the moment of inertia, and $E_S$ is a surface pressure term
(the surface is located at the virial radius $r_{vir}$).
Thus for a DM halo in HE ($\ddot{I}=0$) we expect 
kinetic energy $K_{vt}=(E_S-W)/2$.
Let us identify any variation from this as being 
due to an out of equilibrium state, 
which will also be reflected in turbulence.  
Then we can compute from the DM a measure 
of the degree to which the system is out of equilibrium:
\begin{equation}
\bar{f}_{neq}  = 
\frac{\vert K-K_{vt} \vert}{K} = 
\left\vert \frac{2K}{E_S-W}-1 \right\vert  \;.
\end{equation}
We will assume that the level of nonthermal support
in the gas is proportional to this measure, or
$\bar{f} = \kappa \bar{f}_{neq}$,
where $\kappa$ is a constant independent of mass and redshift.
Thus, limiting
the thermal pressure to be less than the total at
the outermost gas radius $r_f$
(i.e.\ $\alpha(r_f/r_{500})^{n_{nth}}\leq 1$), 
we have
\begin{equation}
\alpha = {\rm MIN}\left[ \; \kappa \bar{f}_{neq}/f^\prime_{pol}, \; 
  (r_{500}/r_f)^{n_{nth}} \; \right] \;.
\label{eq:alpha}
\end{equation}
It only remains to set 
the constant of proportionality $\kappa$; once this is known,
$P_{nth}$ and $P_{th}$ can be computed in the polytropic model.
We will determine $\kappa$ in two ways: by matching the observed
electron pressure at $r_{500}$, and by computing the thermal
pressure directly from the gas in a simulated cluster which
includes full hydrodynamics.

\begin{figure}
\plotone{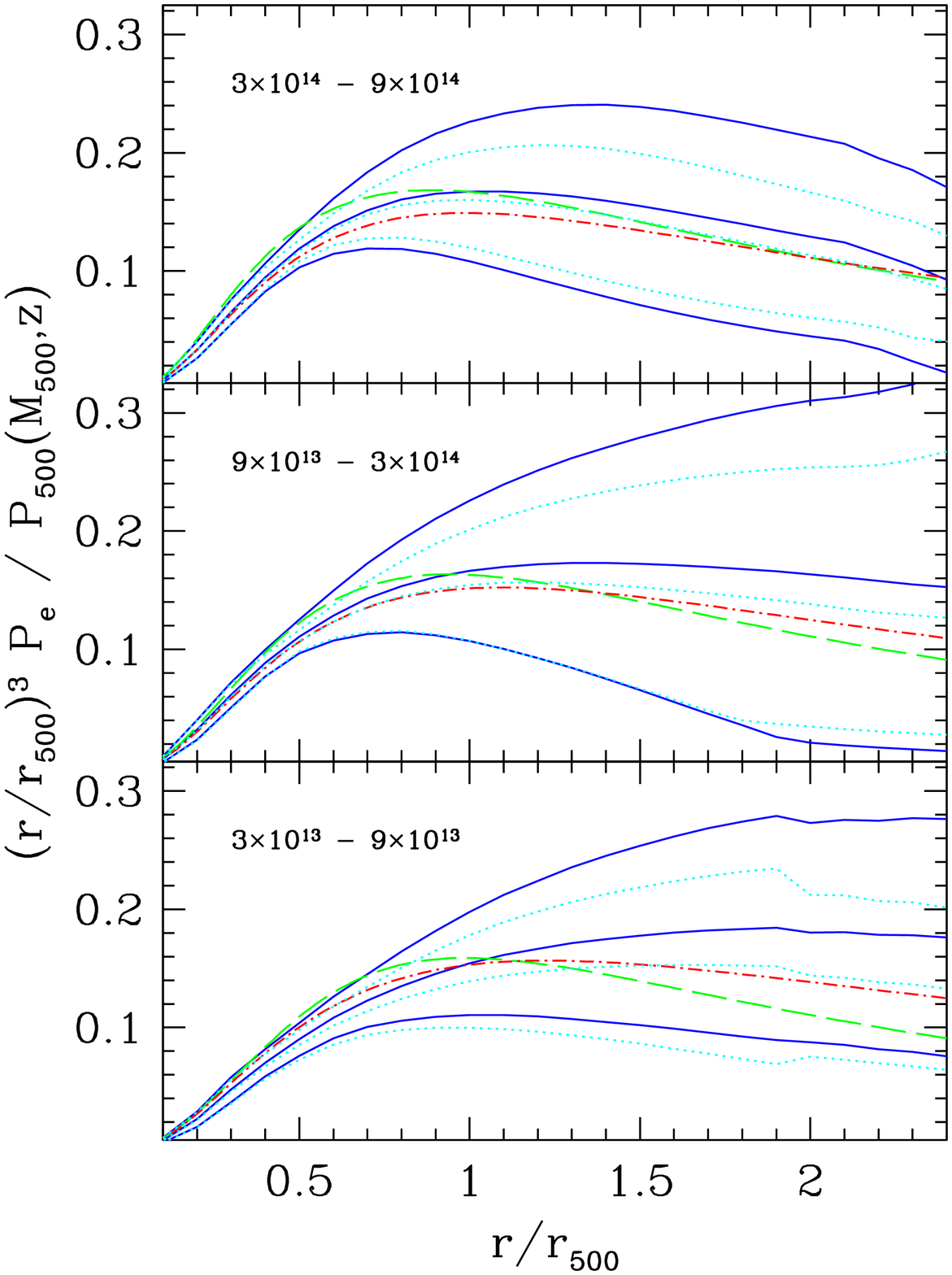}
\caption{Electron pressure profiles of model clusters.
Each panel is labeled by the mass range used; the mean
redshift is $z\approx 0.1$.
The dashed line is the universal profile from X-ray data
by \citet{ArnaudPPBCP10}, who give an expression for
the normalization factor $P_{500}$.
The solid lines indicate the mean and one standard deviation for
the model clusters when basing $\alpha$ on the dynamics
(Eqn.\,\ref{eq:alpha}).  
The dotted lines instead use constant
$\alpha_0=0.18$ in Eqn.\,\ref{eq:zevo},
following \citetalias{ShawNBL10}.
Dot-dashed lines are the mean simulated profile of \citet{BattagliaBPSII},
based on the mean $M_{200}$ and redshift of each bin.
\label{fig:profcomp} }
\end{figure}

From X-ray data,
\citet{ArnaudPPBCP10} derived a electron pressure profile of 
low redshift clusters;
\citet{SunSVDJFVS11} found this same profile holds
for local galaxy groups.
This profile is shown as a dashed line in Fig.\,\ref{fig:profcomp}.
Note that, due to the paucity of observations,
outside of $r_{500}$ this profile is derived from simulations;
also, for higher redshifts it assumes self-similar scaling.
To compare the model profiles with this universal form,
we divided a sample of low redshift ($z<0.2$)
DM halos from the lightcone into three mass bins:
$3\times 10^{13} \leq M_{500}/M_\odot < 9\times 10^{13}$; 
$9\times 10^{13} \leq M_{500}/M_\odot < 3\times 10^{14}$; and 
$3\times 10^{14} \leq M_{500}/M_\odot < 9\times 10^{14}$. 
The mean pressure profile was calculated for each mass bin,
and $\kappa$
was varied to minimize the difference between the observed
and model profiles at $r_{500}$.  
For this purpose $P_{th}$ was converted to electron pressure
and normalized by $P_{500}$ \citep{ArnaudPPBCP10};
the observed profile was calculated using the mean mass and
redshift of the model halos in each bin.
The best fit value is 
$\kappa=0.70$;
the resulting profiles are shown as solid
lines in Fig.\,\ref{fig:profcomp}
(the other parameters in the polytropic model are as
listed in Sec.\,\ref{sec:model}).

Steepening the radial dependence, by
increasing  $n_{nth}$ to as much as 1.2, did not produce
a significantly improved match to the universal profile,
so we leave $n_{nth}$ at the
\citetalias{ShawNBL10}  value of $0.8$.
(It may be that $n_{nth}$ depends on how relaxed a
cluster is; see \citealt{NelsonRSN11z}).
Cutting $\epsilon_F$ in half, to $4\times 10^{-6}$,
also did not change the best fit $\kappa$.
Note
the outermost radius of the polytropic gas, $r_f$, varies from cluster
to cluster. 
In most cases $r_f$ does not extend much beyond $2.6r_{500}$
for the high mass bin;
this limit increases to $3.2r_{500}$ for the middle bin,
while for the lowest mass bin it can extend to $4r_{500}$. 
The gas outside of $r_f$ needs to be treated separately when
making SZ maps \citep{SehgalBDHHLOT10}.

As an alternative method of fixing $\kappa$
(and thus the level of nonthermal pressure),
we compare the resulting profile to that of
a cluster taken from a hydrodynamical simulation \citep[][]{Cen11}
performed with an Eulerian adaptive mesh refinement code, Enzo
\citep[][]{1999aBryan, 1999bBryan, 2004OShea, 2009Joung}.  
First a low resolution simulation with a periodic box 
of $120~h^{-1}$Mpc on a side was run, and a region centered on
a cluster of mass of $\sim 2\times 10^{14}M_\odot$ was
identified.
We then resimulated this subbox with high resolution, but embedded
in the outer $120h^{-1}$Mpc box in order to properly take into account the
large-scale tidal field
and the appropriate boundary conditions at the surface of the refined region.
The subbox centered on the cluster has a size of
$21\times 24\times 20h^{-3}$Mpc$^3$.
The initial conditions in the refined region have a mean interparticle
separation of $117h^{-1}$kpc comoving and a dark matter particle mass
of $1.07\times 10^8h^{-1}M_\odot$.
The refined region is surrounded by two layers of buffer zones,
each of $\sim 1h^{-1}$Mpc,
with particle masses successively larger by a factor of $8$ for each layer;
the outermost buffer then connects with
the outer root grid, which has a dark matter particle mass $8^3$ times
that in the refined region.
The mesh refinement criterion is set such that the resolution is 
always better than $460h^{-1}$pc physical, corresponding to a maximum
mesh refinement level of $11$ at $z=0$.

The cosmological parameters of the simulation are
($\Omega_b, \Omega_m, \Omega_{\Lambda}, h, n_s, \sigma_8$) =
(0.046, 0.28, 0.72, 0.70, 0.96, 0.82).
The simulation includes
a metagalactic UV background
\citep[][]{1996Haardt}
and a model for shielding of UV radiation by neutral hydrogen 
\citep[][]{2005Cen}.
It  also includes metallicity-dependent radiative cooling 
\citep[][]{1995Cen}.
Star particles are created in cells that satisfy a set of criteria for 
star formation proposed by \citet[][]{1992CenOstriker};
star particles typically have masses of $\sim$$10^6M_\odot$.
Supernova feedback from star formation is modeled following \citet[][]{2005Cen}.

\begin{figure}
\plotone{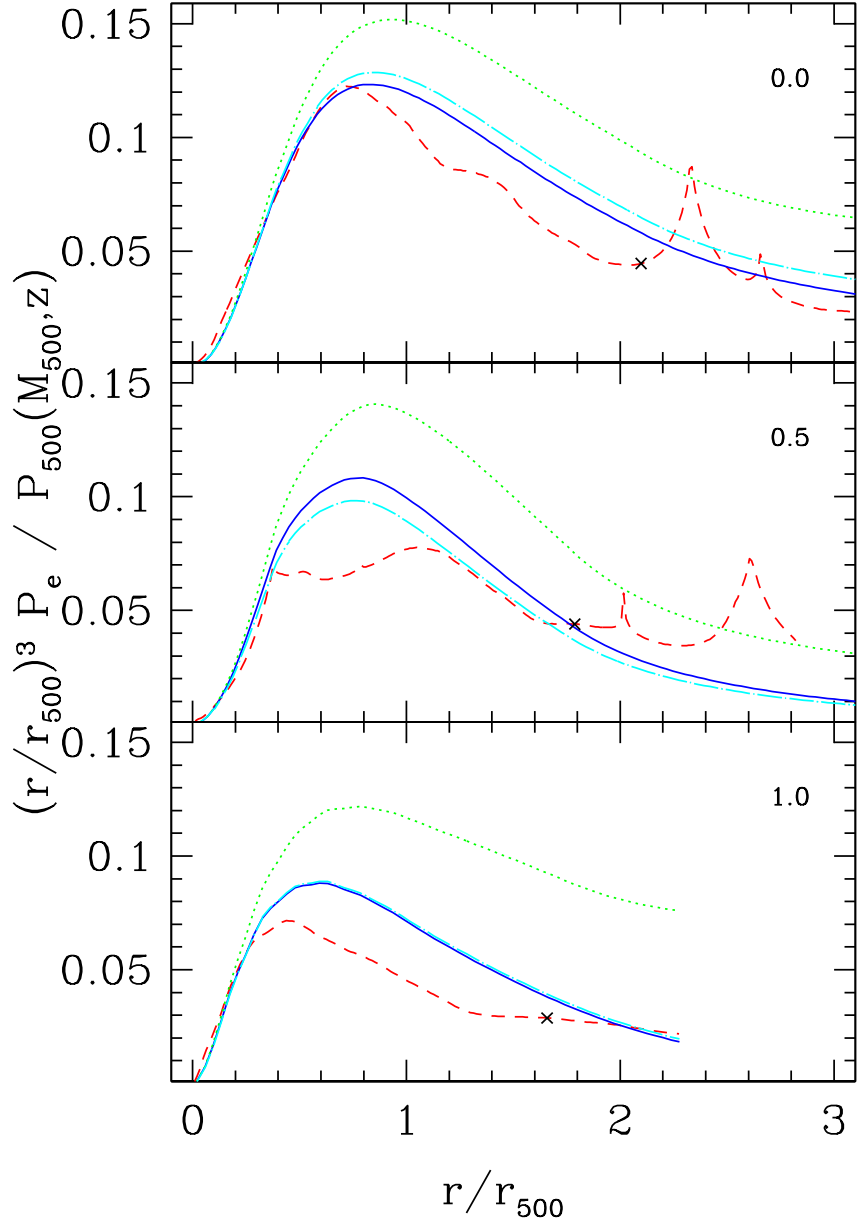}
\caption{Comparison of radial profiles of electron pressure
from a hydrodynamical simulation (red dashed lines) to
the polytropic model applied to the DM potential.
Each panel is labeled with the redshift.  The $\times$ marks
the virial radius.
The dotted line is the model without nonthermal pressure.
The solid line includes it, with $\kappa=0.7$ in Eqn.\,\ref{eq:alpha}.
The dot-dashed line instead uses the $\kappa$
parameter measured directly from the simulation at each redshift.
\label{fig:rcycmp} }
\end{figure}

To emulate the manner in which the polytropic model is normally
applied, the DM particles were taken from the halo at selected
redshifts, and their masses were adjusted 
by a factor $(1+\Omega_b/\Omega_{DM})$
to account for the
baryonic mass.
These particles were then used as input
to the polytropic model code.  The assumed stellar fraction was
increased to match that of the simulation, but feedback was not
included  (i.e.\ $\epsilon_F=0$)
because the simulation did not include AGN.
This cluster grows in mass from
$M_{500}=4.1\times 10^{13}M_\odot$ at $z=1$ to
$M_{500}=1.9\times 10^{14}M_\odot$ at $z=0$.
The thermal pressure profile of the simulated gas at different
redshifts is shown as dashed lines in Fig.\,\ref{fig:rcycmp}.
The polytropic gas
model without turbulent pressure is shown as dotted lines;
this gives a larger pressure in the cluster outskirts than is
seen in the simulated gas.
If we instead fix $\kappa=0.7$, as determined
from the observed pressure profile above, the resulting profiles
are shown as solid lines;
these are much closer to the level of the simulation.  
Alternatively, from the simulated gas one can
compute $\bar{f}$ directly, and thus $\kappa$.  Using the value
of $\kappa$ taken from the simulated gas gives
the dot-dashed curves in Fig.\,\ref{fig:rcycmp}.
At $z=(0,0.5,1)$, the simulated gas gives
$\kappa =(0.57,0.92,0.69)$, for a mean of $0.73\pm 0.18$.
At $z=1.6$, as the cluster is just forming and is still quite unrelaxed,
either choice of parameters gives less turbulent pressure than
is seen in the simulation. 
For comparison with  Eqn.\,\ref{eq:nonthermalpress},
it is also possible to measure $f_{nth}$ as a function of radius
directly from the gas.  A power law with $n_{nth}=0.8$ is a good
fit inside $r_{500}$, but in the outer regions substructures cause
a more variable profile not well represented by a power law.
Also, there are asymmetries; if we measure $f_{nth}$ in 
different octants, there is a variation of roughly 30\%
around the mean at the virial radius.

Based on these two determinations,
in what follows we will fix $\kappa=0.7$ and $n_{nth}=0.8$.
The effect of basing $\alpha$ on the DM halo state in this
manner can be seen by comparing to an
$\alpha$ which varies only with redshift.
We did this by recomputing the profiles using $\alpha$
values from \citetalias{ShawNBL10}, i.e.\ using 
Eqn.\,\ref{eq:zevo} with $\alpha_0=0.18$.
The resulting profiles are
shown as dotted lines in Fig.\,\ref{fig:profcomp}.
We assume a different stellar
fraction at lower masses than did \citetalias{ShawNBL10};
as a result, our mean profile is closer
to the universal profile than they found.
Basing $\alpha$ on the dynamical state of the cluster gives
a larger dispersion in thermal pressure profiles
with a higher mean, particularly at lower masses.
However, even with a constant $\alpha_0$ the dispersion in
$P_{th}$ is large, roughly 50\% near $\sim 2r_{500}$, due
to variations in DM profiles for halos with the same mass.

For a further comparison to simulations,
\citet{BattagliaBPSII} recently derived a fitting function for the mean
$P_{th}$ from a suite of SPH simulations.  For low redshifts, this
mean profile 
(calculated using the mean $M_{200}$ and redshift of the halos
in each bin, and adjusted to electron pressure)
is shown as a dot-dashed line in Fig.\,\ref{fig:profcomp}.
In the outer regions of the halos, this gives profiles similar to those we
found when using the \citetalias{ShawNBL10} form
(Eqn.\,\ref{eq:zevo}) for $\alpha$.
Including mass dependence, as in Eqn.\,\ref{eq:mzevo}, does
not produce a significant change in the mean profile, except
in the lowest mass bin where the mean is 11\% higher near $2r_{500}$.
It is counterintuitive that the mean profile found
when using Eqn.\,\ref{eq:mzevo} is a poorer match to
the fitting function of \citet{BattagliaBPSII} than that found when
using Eqn.\,\ref{eq:zevo}.  The explanation likely lies in the
treatment other physical processes than the nonthermal pressure.
In particular, the amount of star formation has a significant
impact on the entropy of the remaining gas.
Thus differences in how the stellar mass fraction
scales with cluster mass and redshift will affect the
$\rho$ and $P_{th}$ profiles.

\begin{figure}
\plotone{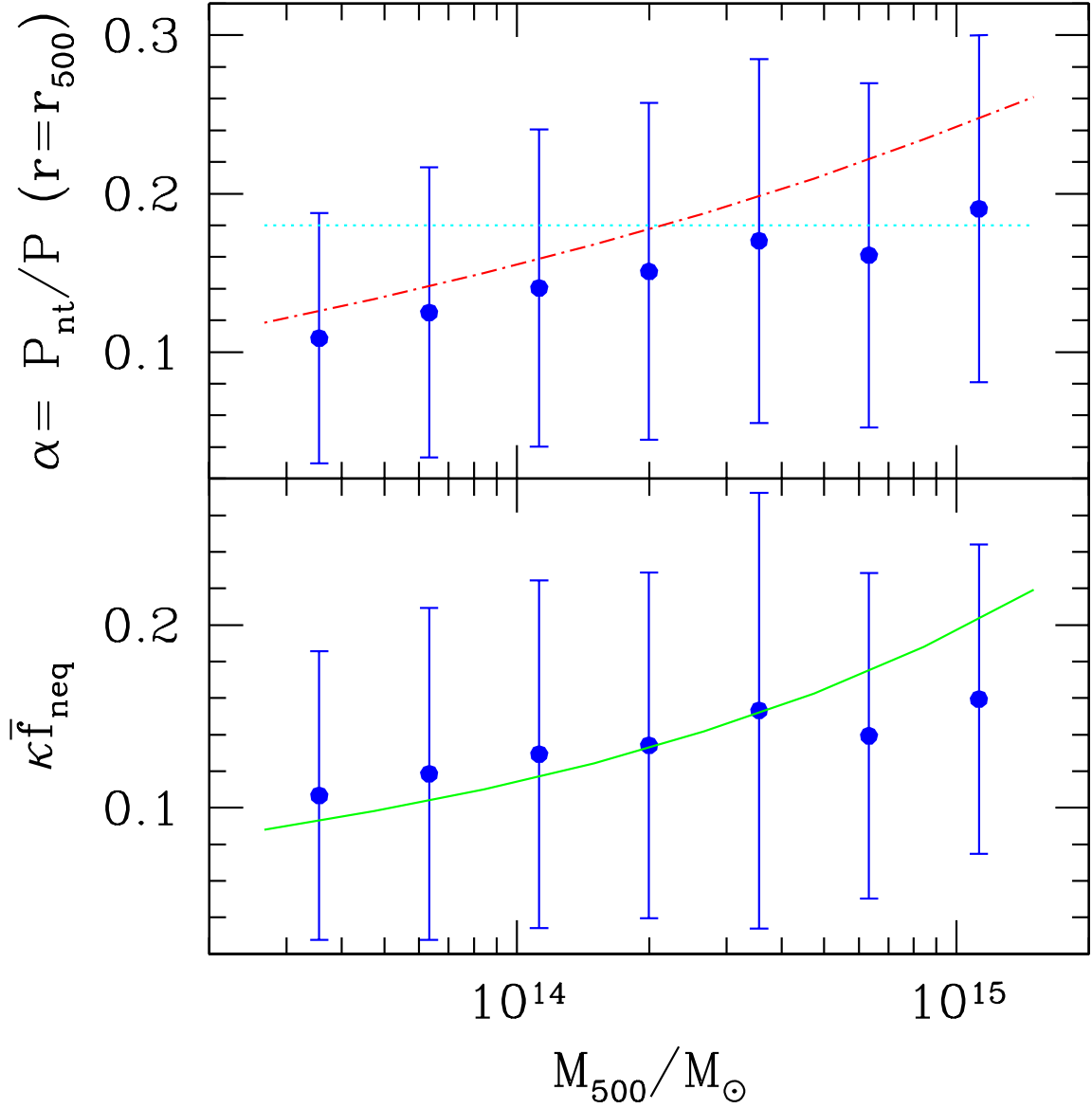}
\caption{ {\it Top:} $\alpha=P_{nth}/P$ at $r=r_{500}$, as a function of mass,
for halos with redshift $<0.2$.
Points with error bars show the mean and standard
deviation in logarithmic mass bins.
The dotted line is the \citetalias{ShawNBL10} value 
at $z=0$, $\alpha_0=0.18\pm0.06$.
The dot-dashed line includes mass dependence as in \citet{BattagliaBPSI}.
{\it Bottom:} $\kappa \bar{f}_{neq}$ of the halos.  The
line is proportional to the peak height $1/\sigma(M)$.
\label{fig:aofm} }
\end{figure}

\begin{figure}
\plotone{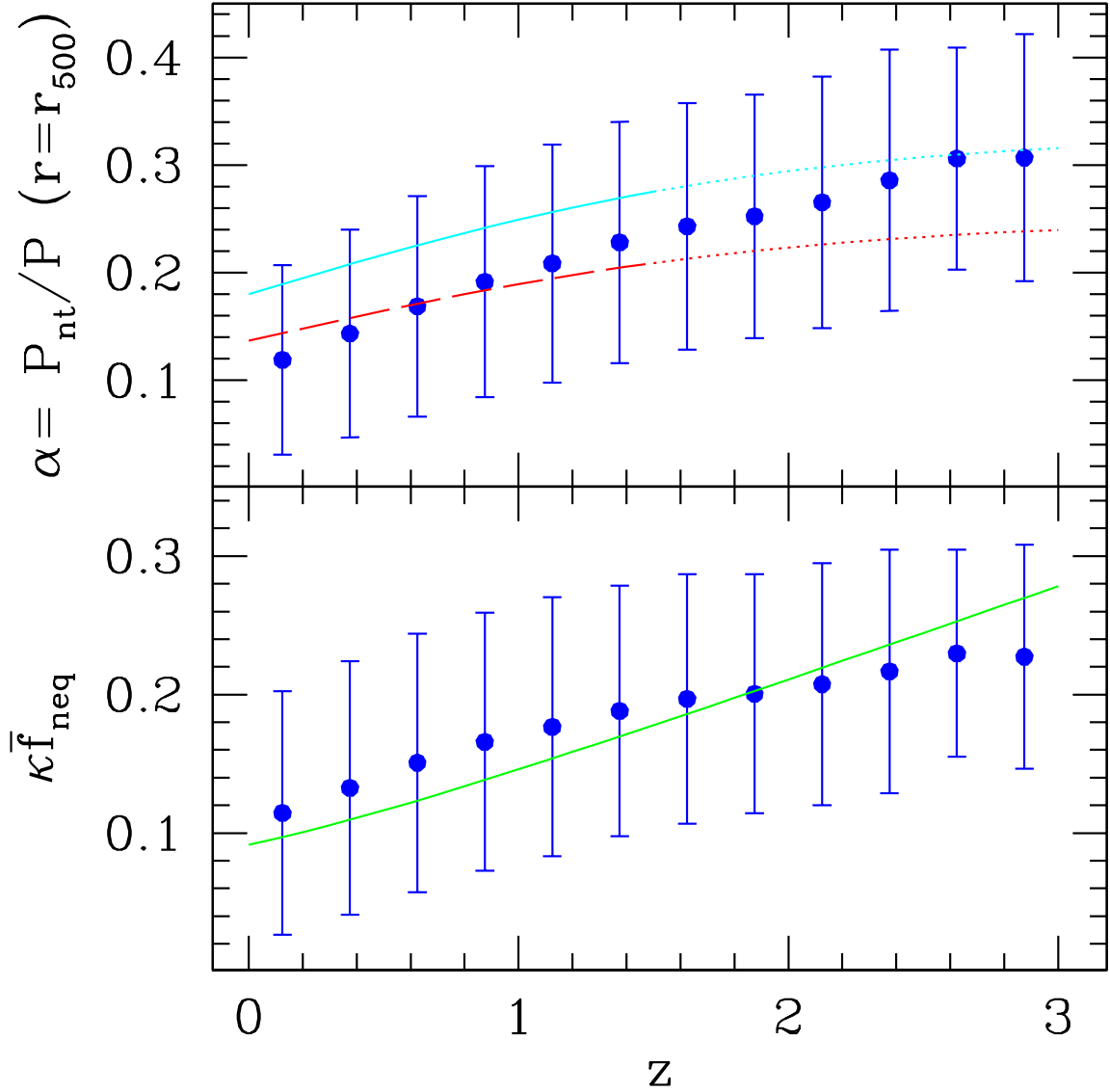}
\caption{ {\it Top:}
$\alpha$ as a function of redshift,
for halos in the range $4\times 10^{13} < M/M_\odot <8\times 10^{13}$.
Points with error bars show the mean and standard
deviation in redshift bins.
The upper line uses the \citetalias{ShawNBL10}
value, $\alpha_0=0.18\pm0.06$;  the lower line
adjusts this by the mass-dependent term of \citet{BattagliaBPSI}.
Note that these latter two fits are based on $z<1.5$ clusters,
and the lines are extrapolated beyond this.
{\it Bottom:} $\kappa \bar{f}_{neq}$ of the halos.  The
line is proportional to the peak height $1/\sigma(M)$.
\label{fig:aofz} }
\end{figure}

The variation of $\alpha$ with mass at low $z$
(using all halos in the light cone with $z<0.2$)
is shown in Fig.\,\ref{fig:aofm}. 
At higher masses, the mean is close to the 
\citetalias{ShawNBL10} value of 0.18, but there is a weak
mass dependence in the same sense as seen by \citet{BattagliaBPSI}.
However, the mean $\alpha$ scales roughly as 
$M_{500}^{0.15}$, a slightly shallower scaling than in Eqn.\,\ref{eq:mzevo}.
The increase is due primarily to an increase in 
$\bar{f}_{neq}$ with mass, which is shown in the
lower panel of Fig.\,\ref{fig:aofz}.   More massive
halos, being rarer, will likely have collapsed more
recently and thus be more out of dynamical equilibrium.
We can quantify this expectation by considering the peak
amplitude $1/\sigma(M)$, where $\sigma(M)$ is the
rms density fluctuation on scales corresponding the the
spherical tophat collapse of halo of mass $M$.
The formation rate of halos at a given mass and redshift
will depend on this parameter \citep[e.g.][]{ShethTormen99},
as halos with higher peak amplitudes will be more extreme
fluctuations.
The solid line in the
lower panel of Fig.\,\ref{fig:aofz} is directly proportional
to the peak amplitude, $\kappa \bar{f}_{neq} = 0.092/\sigma(M)$.
This tracks the mean behavior of the DM halos well, except
at the lowest $\sigma$ (highest $M$).
The other factor determining $\alpha$,
the change in $f^\prime_{pol}$  with mass, is less important.
This change arises because of the changing concentration of
the underlying DM halos. 
$f^\prime_{pol}$ increases with concentration, i.e.\ with
decreasing mass;  but as concentration is quite weakly
dependent upon $M$, the change is slight (less than 20\%).

The variation of $\alpha$ with redshift  is shown in the top panel 
of Fig.\,\ref{fig:aofz}. The halos shown are restricted to a small
mass range so that the evolution shown is unaffected by the
change in the mass function with redshift.
At higher $z$, clusters are typically dynamically younger
and less relaxed, hence $\alpha$ is larger.
The increase in $\alpha$ with redshift
is  slightly more pronounced than that found by \citetalias{ShawNBL10}
and \citet{BattagliaBPSI}; the mean scales as
$(1+z)^{0.76}$ out to $z=3$.
Again the main driver of this evolution is the variation of
$\bar{f}_{neq}$, shown in the lower panel.  Assuming  
$\bar{f}_{neq}$ is proportional to peak height reproduces
the trend seen out to $z\approx 1.5$, but again as $\sigma$
becomes small (i.e.\ higher $z$) this leads to an overestimate.
Also, once again the variation of $f^\prime_{pol}$ is smaller,
at about 20\% over the whole redshift range shown.

\begin{figure}
\plotone{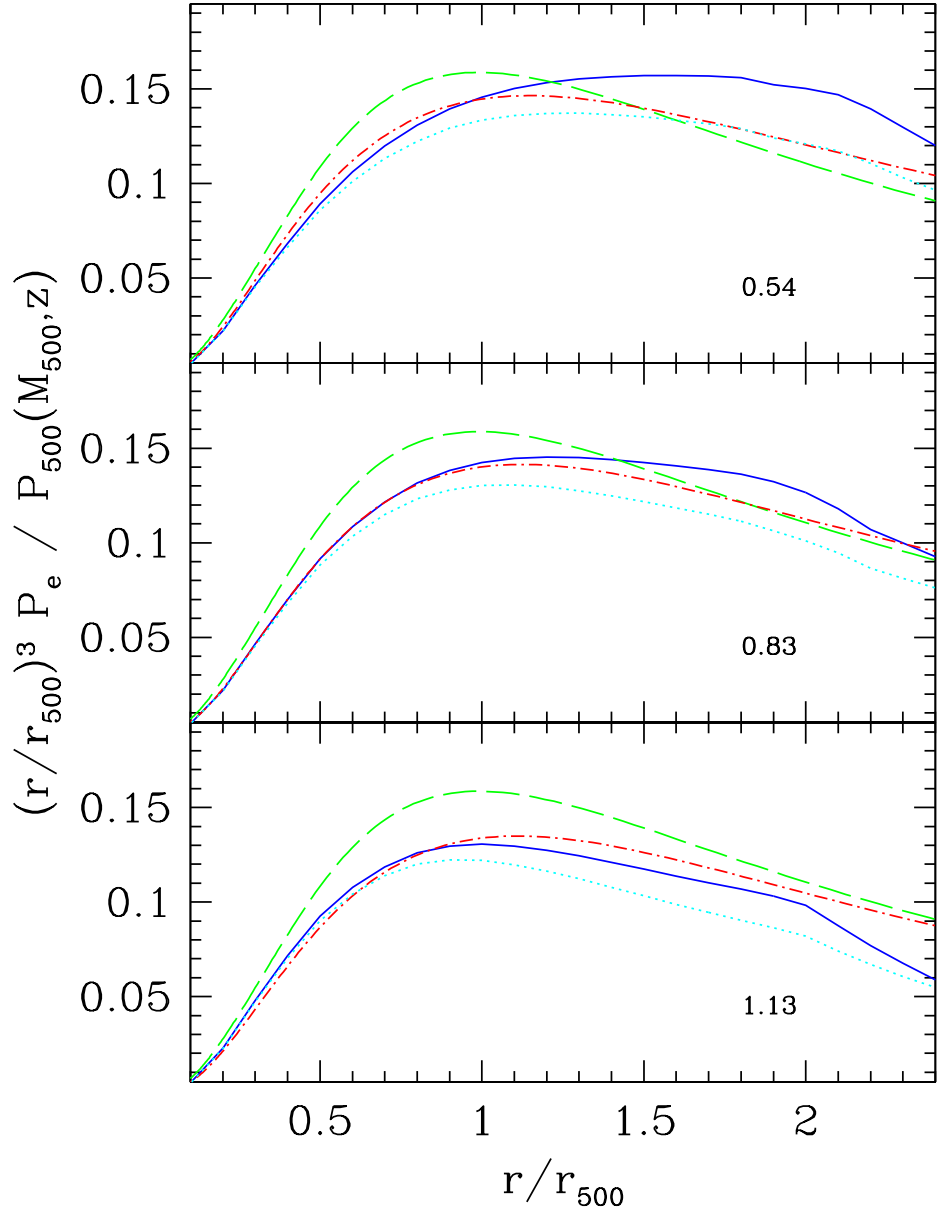}
\caption{Electron pressure profiles at different redshift.
Solid lines: the mean adopted model profile, for selected halos in
the mass range 
$3\times 10^{13} < M_{500}/M_\odot < 9\times 10^{13}$ and
in redshift bins $0.4<z<0.7$, $0.7<z<1.0$, and $1.0<z<1.3$
(panels are labeled with mean redshift).  
Dashed lines:  the low-redshift profile of 
\citet{ArnaudPPBCP10}, assuming self-similar scaling to higher $z$.
Dot-dashed lines:  the mean simulated profile of \citet{BattagliaBPSII}.
Dotted lines: mean profile using Eqn.\,\ref{eq:zevo} for $\alpha$
\citepalias{ShawNBL10}.
\label{fig:profvz} }
\end{figure}

The evolution with redshift of the mean pressure profile
is shown in  Fig.\,\ref{fig:profvz}.
A lower mass range was chosen so that the mean mass
$\approx 5.6\times 10^{13}M_\odot$ does not change with $z$.
The model profiles are compared to the 
\citet{ArnaudPPBCP10} profile, which assumes a self-similar evolution.
Our model deviates from this assumption; the fit of 
\citet{BattagliaBPSII} deviates in the same manner, but 
not as strongly.  The main processes which can cause a break
from self-similarity are star formation and feedback; without
matching the evolution of these processes it is difficult
to compare methods.  However, 
we have already seen that
our method does give a
stronger scaling of $\alpha$ with redshift
(cf. Fig.\,\ref{fig:aofz}).

\subsection{Setting the level of feedback} \label{sec:feedback}

The level of feedback, parametrized as $\epsilon_{F}M_*c^2$,
needs to be determined.
To constrain $\epsilon_{F}$, we fit to X-ray observations in
the $kT-E(z)M_g$ plane, where
$E^2(z)=\Omega_m(1+z)^3+\Omega_{\Lambda}$. 
The two samples we compare to are the cluster sample
of \citet{MantzAERD10} and the lower mass sample of
\citet{SunVDJFV09}.
$M_g$ is found inside $r_{500}$,
and $T$ is the spectroscopic temperature inside the radial
range $[0.15-1]r_{500}$.  The simulated sample includes
all halos in the light cone with $M_{500}>5.5\times 10^{14}M_\odot$
and $z<0.45$, 79 in all, plus another 79 chosen from $z<0.1$
to span the range $2\times 10^{13} < M_{500} < 1.5\times 10^{14}$.

\begin{figure}
\plotone{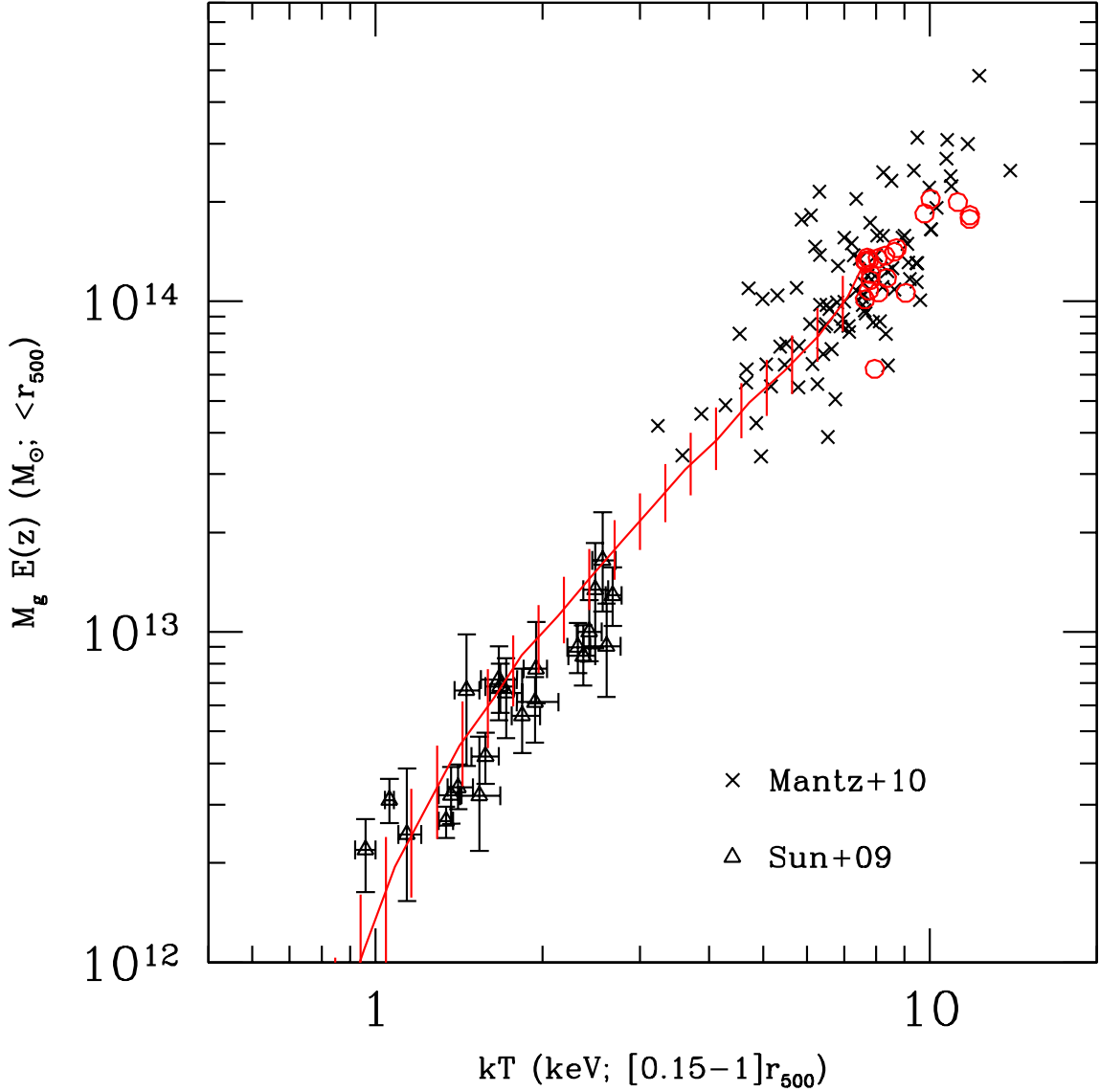}
\caption{ Comparison of the model to X-ray observations by
\citet{SunVDJFV09} and \citet{MantzAERD10}; for the latter,
error bars have been omitted for clarity.
The red line is the median and vertical bars enclose 68\% of the model halos;
at the high $T$ end individual halos are shown as circles.
\label{fig:mtmgfig} }
\end{figure}

To find the best fit we minimize  $\chi^2$, defined as follows.
For simulated halo $i$ and observed cluster $j$
(with observational uncertainties $\sigma_M$ and $\sigma_T$),
\begin{mathletters}
\begin{eqnarray}
c_{ij} = \left( \frac{T_i-T_j}{\sigma_{Tj}} \right)^2 + 
         \left( \frac{M_i-M_j}{\sigma_{Mj}} \right)^2      \;, \\
c_i = \left( \sum\limits_{j} [ c_{ij} ]^{-1} \right)^{-1} \;, \\
\chi^2 = \sum\limits_{i} c_i \;.
\end{eqnarray}
\end{mathletters}
The best fit feedback level 
when including our model for $\alpha$
is $\epsilon_F=8\times 10^{-6}$.  
If we instead use Eqn.\,\ref{eq:zevo},
the same value is recovered.
The X-ray properties of the model are compared to the observed
samples in Fig.\,\ref{fig:mtmgfig}.

This amount of feedback is larger than that in the Standard model
of \citetalias{BodeOV09}, which has no nonthermal pressure.
The Standard model value was also chosen to match observed clusters
in the $kT-E(z)M_g$ plane.  
Starting with this model,
the addition of nonthermal
pressure reduces $T$ without making any change to $M_g$,
so at a given $T$ the value of $M_g$ is now too high.
Increasing the amount of feedback both reduces $M_g$
and increases $T$, bringing the model back into agreement
with the observed relation.
This amount of
feedback is also larger than used by \citetalias{ShawNBL10},
but we are also using a lower amount of star formation.
As discussed in \citetalias{BodeOV09}, the star formation and feedback
parameters are somewhat degenerate, such that increasing the
amount of star formation in lower mass clusters would require
a lower feedback parameter.
Thus a greater amount of feedback is to be expected.

As mentioned in Sec.\,\ref{sec:model},
it is possible to also include
relativistic pressure, parametrized as 
$\delta_{rel}$ times the non-relativistic pressure. However,
repeating the above procedure while varying both 
$\epsilon_F$ and $\delta_{rel}$ yields
a best fit with $\delta_{rel}=0$, so we do not include it. 
The main effect from including $\delta_{rel}$ is to increase
the total nonthermal pressure in the core.  However, there are
other, compensating effects we are not including, such as
cooling and the presence of a cD galaxy.

\section{Implications for the SZ signal} \label{sec:discuss}

\begin{figure}
\plottwo{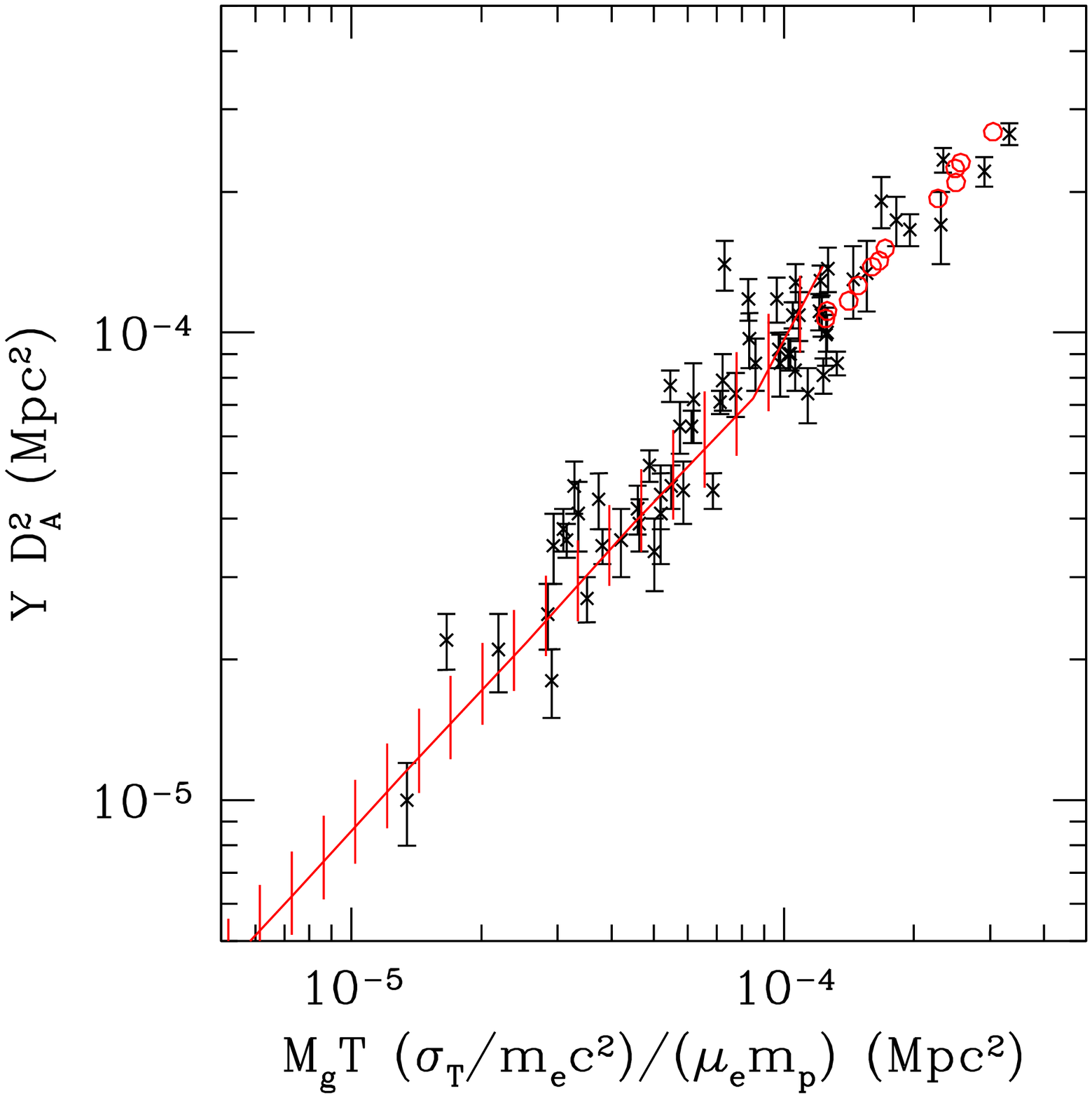}{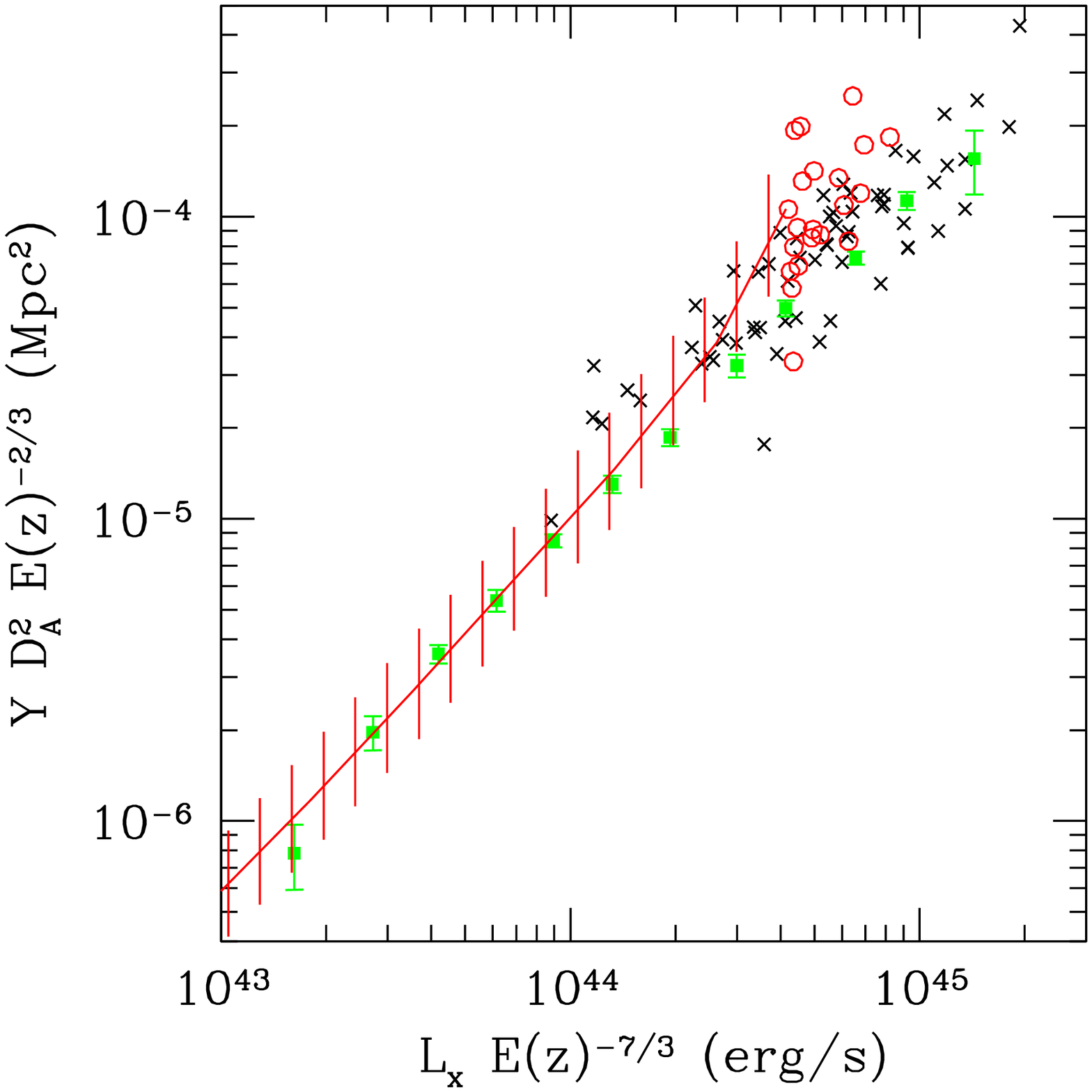}
\caption{  Comparison of the model to X-ray and SZ observations by
\citet{PlanckXI11,PlanckX11}.
Left: points with error bars are from \citet{PlanckXI11}.
The red line is the median and vertical bars enclose 68\% of the model halos;
at large $Y$ individual halos are shown as circles.
$M_g$ is measured inside $r_{500}$ and $T$
in the radial range [0.15-0.75]$r_{500}$. 
All model halos in the light cone with $z<0.5$ are included.
Right: SZ signal versus X-ray luminosity in the band
[0.1-2.4]keV, measured inside $r_{500}$.  Squares with error
bars are binned clusters from \citet{PlanckX11}; crosses
are individual clusters from \citet{PlanckXI11} with error
bars omitted for clarity.
\label{fig:pxyfig} }
\end{figure}

Having set the parameters of the model by comparison to
X-ray observables and simulations, it is now possible to
make predictions concerning the SZ signal.
We begin by comparing to early results from the Planck satellite. 
\citet{PlanckXI11} combine measurements of 
$D^2_AY=\int P_{th}dV$ inside $r_{500}$ 
with X-ray data taken by XMM-Newton for 62 $z<0.5$ clusters.
$Y$ is expected to scale closely with the X-ray quantity $Y_X=CM_gT$.
Here $C=\sigma_T/(m_ec^2\mu_em_p)$, where $\sigma_T$ is the
Thompson cross-section, $m_ec^2$ is the electron rest mass,
and $\mu_em_p$ is the mass per electron in a fully ionized plasma.
The data from \citet{PlanckXI11} is shown 
in the left-hand panel of Fig.\,\ref{fig:pxyfig};
the gas mass $M_g$ is measured inside $r_{500}$, and $T$
is measured in the radial range [0.15-0.75]$r_{500}$. 
For the $f_{nth}$ model clusters we include
all halos in the light cone with $z<0.5$.
Fitting those with $D^2_AY>10^{-5}$Mpc$^2$, the slope of the
$D^2_AY-Y_X$ relation is consistent with unity;
this agrees with \citet{PlanckXI11}, but is roughly 2-sigma
steeper than the slope found by \citet{RozoVM12z}.
If, following \citet{RozoVM12z}, we make the pivot point
$8\times 10^{-5}$Mpc$^2$, then the model gives
$\ln(D^2_AY/Y_X)=-0.152\pm0.002$ with rms scatter $\sigma=0.031$;
in other words $(D^2_AY/Y_X=0.86$, significantly lower than
the best fit given by \citet{PlanckXI11}.
On the other hand, it is higher than the analysis of a subsample
of the Planck clusters using different X-ray data by
\citet{RozoVM12z}, who find $D^2_AY/Y_X=0.82\pm0.024$.
These results are insensitive to the redshift range used,
but including clusters with $D^2_AY$ below $10^{-5}$Mpc$^2$
lowers the ratio to $(D^2_AY/Y_X=0.82$ and increases
the scatter to $\sigma=0.037$.

The relation between SZ signal and the X-ray luminosity
is shown in the right-hand panel of Fig.\,\ref{fig:pxyfig}.
The luminosity $L_X$ is in the band [0.1-2.4]keV, measured inside $r_{500}$. 
The same clusters from \citet{PlanckXI11} are shown,
as well as data for lower-mass clusters from
\citet{PlanckX11}, who measured $Y$ for roughly 1600
X-ray selected clusters and  averaged the results
in bins of $L_X$.
The model reproduces the observed relation well
except at the highest $L_X$, where the model cores
are not dense enough.
$Y_X$ depends roughly linearly on gas density, and the core was excluded
when computing $T$.   In contrast, $L_X$ depends on the square of
density, and thus strongly weights the cluster center.
Our model does not include cooling in the cluster core;
such cooling would result in higher $L_X$ without
significantly affecting $Y$ or $Y_X$.
The slope of the $f_{nth}$ model $L_X-Y$ 
relation  (set mostly by the lower mass clusters)
is 1.09, in good agreement with
\citet{PlanckX11}.
For this model,
the evolution of $Y$ with redshift follows the standard
self-similar scaling out to $z=1$.

\begin{figure}
\plotone{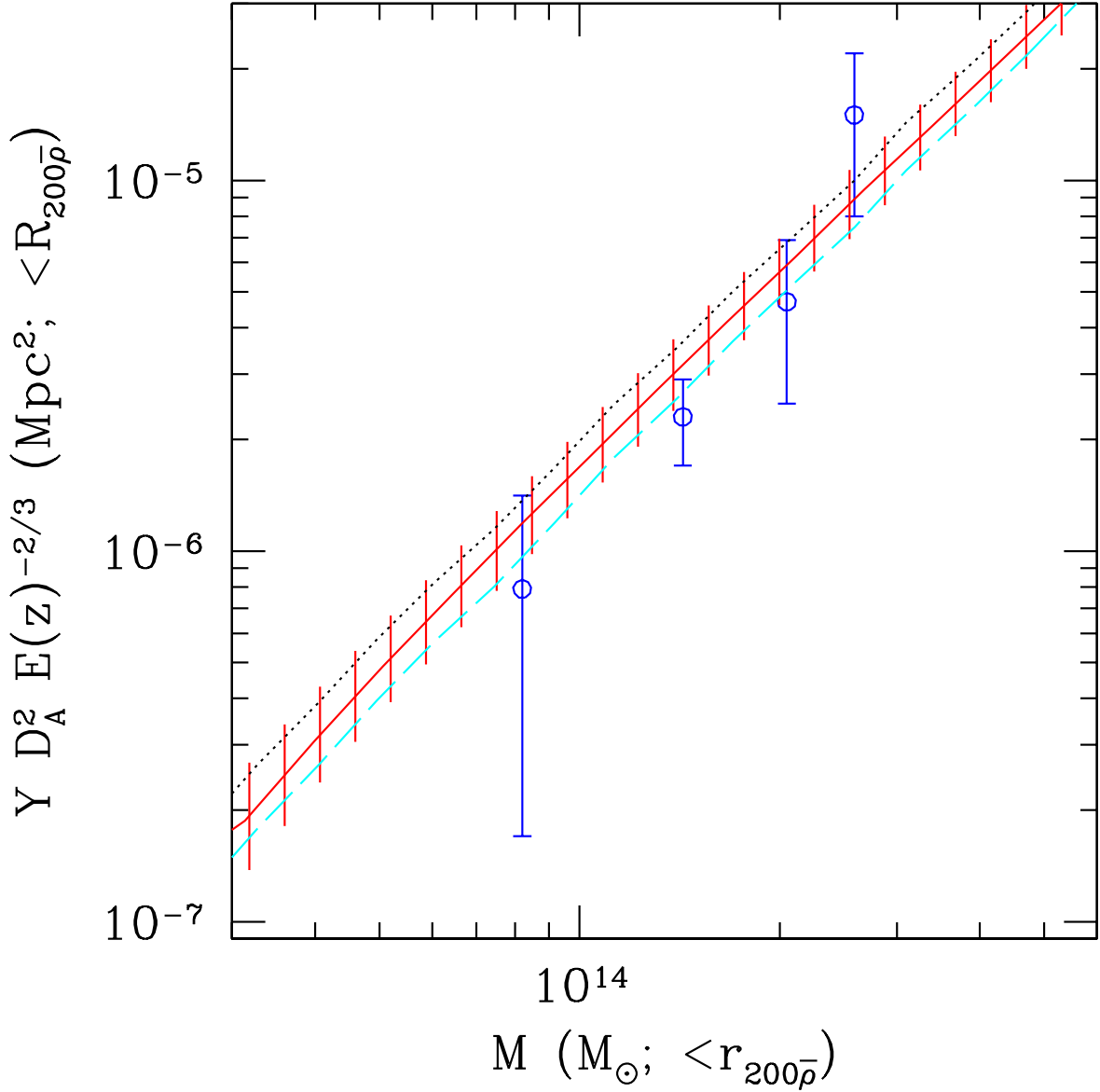}
\caption{
Projected cylindrical SZ decrement for halos hosting LRGs.
Circles with error bars are from ACT data, with
masses estimated from weak lensing \citep{Hand11}.
The radio-quiet LRG sample is shown; masses are measured
inside 200 times the mean density.
The solid red line with vertical bars is 
the median and 68\% range of the $f_{nth}$ model clusters.
The dotted line is the median for the Standard model of
\citetalias{BodeOV09}, and the dashed is for the nonthermal20
model of \citet{TracBO11}.
\label{fig:handcomp} }
\end{figure}

The above comparisons were all measured within of $r_{500}$,
whereas the effect of nonthermal pressure is greatest
outside this radius.  Recently
\citet{Hand11} used ACT data to measure the $Y$
out to a cylindrical radius corresponding to a cluster
density of 200 times the mean.
This includes the entire contribution along the line of sight,
as well as a larger solid angle than subtended by $r_{500}$,
and so is more sensitive to the cluster outskirts.
The measurement was done by stacking the locations of LRGs,
which reside in massive halos, to find the
projected SZ signal.
This data is shown in Fig.\,\ref{fig:handcomp}.
Only the radio-quiet LRG sample is shown, to avoid contamination.
The masses of the halos are determined from weak lensing,
as the masses determined from bias may be affected by 
systematics \citep{More11}.
The model fits the data  within the errors.

Fig.\,\ref{fig:handcomp} also shows the 
mean relation for the
Standard model of \citetalias{BodeOV09}.
This model has no nonthermal pressure and a lower amount
of feedback, such that inside $r_{500}$ it would closely
resemble the  $f_{nth}$  model.  Because of higher thermal 
pressure outside this radius, it shows a larger SZ signal.
Including $f_{nth}$ has little
effect on the slope of the relation, but reduces $Y$
at a given mass by 15\% from the Standard model.
We also show the nonthermal20 model of \citet{TracBO11}; 
this has constant 20\% nonthermal pressure at all radii
(note it has a different
level of star formation and feedback as well).
Thus, while it has a higher thermal pressure at large radii,
it will have more nonthermal support in the core, and thus
the central density and thermal pressure will be lower.
This leads to a smaller value of $Y$ at a given mass.
However, the slope of this relation is again little changed.

The reduced SZ signal predicted when $f_{nth}$ is included will
alter the predicted angular power spectrum in SZ maps.
Computing the $C_l$ in the same manner as 
\citet{TracBO11}, the $f_{nth}$ model yields
a thermal SZ template 24\% lower than the Standard model
(at 280 GHz, for $\sigma_8=0.8$).
Two effects which one might expect could
alter the shape of the template, the
generally lower $P_{nth}$ in cluster outskirts and the variation in
the level of $f_{nth}$ from cluster to cluster,
appear not to do so;  the peak of the SZ power, at $l\approx 3000$,
remains the same. 
This is in contrast to the nonthermal20 model, which is 45\% lower
than the Standard model, and has a peak at slightly larger angular
scales.  The reduction of 24\% from the Standard model used
in \citet{SehgalBDHHLOT10} will alleviate somewhat, but not
eliminate, the tension with observations 
\citep{SehgalBDHHLOT10,DunkleyACT11,ReichardtSPT11z}.

To summarize,
in this paper we have presented a method of including turbulent
pressure in a model of the ICM, similar to that of \citetalias{ShawNBL10}.
The fraction of pressure in turbulent form is allowed to vary from
cluster to cluster, depending on the host halo's dynamical state.
The normalization is set by comparing the pressure profiles to X-ray
observations and hydrodynamical simulations.
The nonthermal pressure fraction measured at $r_{500}$
is typically in the range 10-20\%, trending higher with cluster
mass and with redshift.
This will be useful in creating improved templates for interpreting
new data from millimeter and microwave instruments such as the PLANCK
satellite.  However, a better understanding of the thermodynamic
state of the ICM at higher redshifts will be needed to fully
exploit this new data.

\acknowledgments
This work was supported by National Science Foundation  grant 0707731.
Computer simulations and analysis were supported by the
NSF through resources provided by XSEDE,
the Pittsburgh Supercomputing Center,
and the National Center for Supercomputing Applications
under grant AST070015; computations were also
performed at the TIGRESS high performance
computer center at Princeton University, which is jointly supported by
the Princeton Institute for Computational Science and Engineering and
the Princeton University Office of Information Technology.

We would like to thank Dr. M.K.R.\ Joung for help on
generating initial conditions for the simulations and for running a portion
of them, and Greg Bryan for his help with the Enzo code.
This work was supported by NSF grant
0707731 and NASA grant NNX11AI23G.
Computer simulations and analysis were supported by the
NSF through resources provided by XSEDE,
the Pittsburgh Supercomputing Center,
and the National Center for Supercomputing Applications
under grant AST070015;
computing resources were also provided by the NASA High-
End Computing (HEC) Program through the NASA Advanced
Supercomputing (NAS) Division at Ames Research Center.
Computations were also
performed at the TIGRESS high performance
computer center at Princeton University, which is jointly supported by
the Princeton Institute for Computational Science and Engineering and
the Princeton University Office of Information Technology.

\bibliographystyle{hapj}
\bibliography{ms}

\epsscale{0.8}

\epsscale{1.0}

\epsscale{0.8}
\epsscale{1.0}

\end{document}